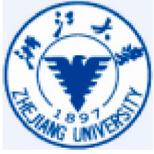
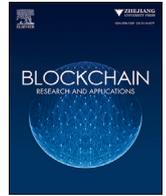

Research Article

# Integrating blockchain technology within an information ecosystem

Francesco Salzano [a,*], Lodovica Marchesi [b,*], Remo Pareschi [a], Roberto Tonelli [b]

[a] *Stake Lab, University of Molise, Campobasso 86100, Italy*
[b] *Dep. of Mathematics and Computer Science, University of Cagliari, Cagliari 09100, Italy*

A R T I C L E   I N F O

A B S T R A C T

**Context:** Blockchain-based information ecosystems (BBIEs) are a type of information ecosystem in which blockchain technology is used to provide a trust mechanism among parties and to manage shared business logic, breaking the traditional scheme of information ecosystems dominated by a leading company and leveraging the decentralization of data management, information flow, and business logic.
**Objective:** In this paper, we'd like to propose an architecture and the technical aspects concerning creating a BBIE, underlining the supplied advantages and the logic decomposition among the business and storage components.
**Method:** The requirements are derived from the current needs of the collaborative business and the data collected by surveying practitioners. To meet these needs, we followed the Grounded Theory research approach. We validate our architectural schema against a case study on managing a wine supply chain involving different companies and supervision authorities.
**Results:** The proposed solution integrates blockchain-based applications with the existing information system as a module of the ecosystem, leveraging on the low costs, scalability, and high-level security because of the restricted access to the network.
**Conclusion:** We must go a long way in deepening and refining the possibilities offered by technology in supporting innovative multi-organizational business models. BBIEs can contribute substantially to paving the way in such a direction.

## 1. Introduction

The latter half of the 20th century saw the evolution of companies closely intertwined with the advancement of information systems [1,2]. An information system integrates technologies, processes, and human collaboration to produce data that help organizations meet their strategic objectives. This concept has evolved significantly and was shaped by technological advances, shifting business landscapes, and emerging theories on enhancing decision-making and problem-solving through better information management [3].

Historically, information systems date back to the 1950s and 1960s, marked by the use of mainframe computers to automate basic organizational functions such as payroll and accounting. The advent of the Internet, the World Wide Web, mobile computing, and cloud technologies has transformed these systems into complex information ecosystems that span multiple business entities [4]. Prominent examples like Amazon and Walmart demonstrate how these ecosystems manage the flow of goods, information, and materials across extensive supply chains, although decision-making often remains centralized within a dominant entity. This centralization highlights the need for robust systems that can adapt to the dynamic requirements of modern enterprises.

Recent studies have emphasized the potential of blockchain and distributed ledger technologies (DLTs) to create fairer, more democratic ecosystems by automating trust and decentralizing decision-making processes [5,6]. Unlike traditional models where a single entity controls trust relationships, DLTs support a trustless environment, enabling equitable power distribution and decision-making among all stakeholders. Furthermore, the distributed architecture of DLTs and blockchain allows for shared business logic and the use of smart contracts (SCs) for autonomous operations [7].

This paper aims to outline a comprehensive architectural framework designed to address the evolving needs of modern businesses characterized by collaborative data sharing and the necessity for real-time supply chain visibility. We propose a high-level architecture for blockchain-based information ecosystems (BBIEs), which incorporates both blockchain and non-blockchain components and is shared among various entities. This architecture is informed by a survey of practitioners across multiple industries to understand their specific requirements.

A BBIE, as explored in prior literature [8], establishes blockchain technology as the foundation for trust and shared business logic, challenging the conventional information ecosystem paradigm dominated by single entities. This concept promotes a framework where trust is inherent and governance is shared across inter-organizational domains.

In BBIEs, decision-making and information flow are collective efforts supported by consensus mechanisms [9]. These systems process transactions through a decentralized network of nodes, enhancing transparency and security while minimizing points of failure.

The emergence of blockchain-as-a-service (BaaS) platforms plays a crucial role in facilitating the adoption of BBIEs [10,11]. BaaS offers cloud-based solutions that allow organizations to develop, host, and deploy blockchain applications and SCs without extensive infrastructure investments or specialized technical skills. This shift promotes the seamless integration of decentralized applications (dApps) with existing information systems, supporting scalable ecosystem implementations.

Moreover, addressing challenges related to interoperability and integration with legacy systems is essential for transitioning towards decentralized architectures [12]. Components such as off-chain data storage and identity management systems (IMSs) are integral to efficiently managing identities across various system layers and enhancing the functionality of blockchain-based information ecosystems.

To validate the proposed architectural framework, we conduct a case study focusing on an agrifood supply chain that includes multiple companies and regulatory bodies. This case study demonstrates the practical applicability and effectiveness of our architectural approach in addressing real-world challenges.

In summary, the primary contributions of this paper are:

1. Defining "BBIEs" as systems where blockchain serves as a trust mechanism and manages shared business logic, facilitating the decentralization of data and information flow.
2. Surveying 11 companies to identify the needs and challenges within business-to-business (B2B) and business-to-customer (B2C) services.
3. Proposing an architectural framework for BBIEs that highlights the advantages and logical decomposition of business and storage components.
4. Evaluating this framework through a case study on agrifood supply chain management, involving various stakeholders and regulatory authorities.

The remainder of the paper is organized as follows. Section 2 presents the background; Section 3 provides a review of existing works; in Section 4, we present how we design and conduct our research using a questionnaire; Section 5 describes the proposed solution, highlighting the decomposition between off-chain and on-chain components; Section 6 shows a case study for a BBIE involving different companies to better explain and validate our solution. Finally, Sections 7 and 8 discuss the advantages and limitations of the proposed architecture and conclude the article.

## 2. Background

This section provides relevant background information to understand the present study and the technologies involved.

### 2.1. Information ecosystem

Information ecosystems encompass a broad spectrum of configurations, from centralized systems controlled by a single entity to decentralized, collaborative networks. Unlike traditional information systems that often center around centralized management, information ecosystems can also be highly decentralized. These ecosystems are comprised of interconnected entities, including businesses, organizations, and individuals, collaborating to achieve shared objectives. In centralized systems like those of Amazon and Walmart, decision-making and control are concentrated. In contrast, decentralized ecosystems distribute authority and decision-making across multiple stakeholders, enhancing transparency, fairness, and inclusivity. Technologies such as blockchain and DLT enable these decentralized systems by establishing robust trust mechanisms and facilitating shared business logic among participants, thus redefining the management, sharing, and utilization of information across diverse administrative domains.

### 2.2. DLT, blockchain, smart contracts, and dApps

DLT represents a fundamental innovation in information technology, with the potential to reshape organizational structures, collaboration, and operations across numerous sectors [13]. DLT systems maintain data across a network of distributed nodes, where the ledger's state is updated through consensus algorithms that reflect the network's level of decentralization.

Blockchain, a type of DLT, was popularized by the introduction of Bitcoin by Satoshi Nakamoto [14] and has since expanded its applications beyond cryptocurrency [15]. Blockchains operate as replicated databases with data stored in a chain of blocks, secured through consensus mechanisms [9,16]. The immutability of blockchains ensures that data, once entered, cannot be altered, with each block linked to the previous one via cryptographic hashes.

SCs automate transaction execution on blockchain networks when predetermined conditions are met, thereby facilitating decentralized decision-making and operations [17,18]. These contracts enable the development of dApps, which are software platforms that operate on blockchain networks, enhancing system transparency, security, and efficiency [6].

### 2.3. Types of DLTs

DLTs vary in their architectural design and operational attributes:

**Public (Permissionless) Blockchains**: open to all, these blockchains embody decentralization and transparency, allowing any participant to view and verify transactions. Examples include Bitcoin and Ethereum, which support an open, global environment for transactions and SCs deployment.

**Hybrid (Public-Permissioned) Blockchains**: these blockchains blend public accessibility with controlled participation. While the ledger is publicly viewable, actions such as transaction validation or SC deployment require specific permissions. Ripple (XRP Ledger) and Stellar exemplify this model, combining the efficiency of permissioned systems with the openness of public blockchains.

**Private and Permissioned Blockchains**: restricted to pre-approved participants, these blockchains are ideal for applications requiring high levels of privacy and control, which are often used in corporate consortia. Private blockchains are typically restricted to a single organization and offer centralized governance, while permissioned blockchains allow for participation from multiple organizations and employ distributed governance models. Both types of blockchains prioritize access control and data privacy but cater to different use cases and requirements. Private blockchains are commonly used for internal business processes within organizations, where data privacy, control, and efficiency are paramount. Permissioned blockchains are suitable for use cases that involve multiple organizations or stakeholders who need to collaborate while maintaining control over access and governance. Hyperledger Fabric exemplifies a platform enabling the development of private blockchain networks tailored to enterprises' needs [5].

**Directed Acyclic Graph-based (DAG) DLTs**: as an alternative to traditional blockchains, DAG-based systems allow for parallel transaction processing, increasing scalability and efficiency. These are particularly suitable for high-volume environments like the IoT, where systems like

IOTA and Nano provide rapid, cost-free transactions ideal for real-time data handling [19,20].

## 3. Related work

### 3.1. Blockchain technology: applications and impacts

Lu and Xu [21] presented *originChain*, one of the first blockchain-based traceability systems built by replacing the centralized database with a blockchain achieving transparent tamper-proof traceability data, and enhancing the data's availability.

Schneider et al. [22] developed a theoretical framework to examine the impact of blockchain on ecosystems and business models. Their research show that blockchain is a combination of technology and human actors from various organizations and society as a whole, who collaborate to create a unique form of agency different from that of humans or machines.

Blockchain technology allows businesses to modify and develop new business models. A recent study by Taherdoost and Madanchian [23] examined the current state of new blockchain-based business models. Their research highlights how blockchain could benefit companies, including cost saving via faster transaction times, disintermediation, less record-keeping than customers due to DLT, and better data tracking and verification. Furthermore, blockchain technology improves the interactions of a company with its customers, competitors, and suppliers.

Blockchain technology was first used in crypto-finance, but now businesses use it to share digital data among participants. It can provide new capabilities by changing the way participants interact with digital transactions [24]. Business operations can benefit from the use of blockchain technology. This includes simplified disintermediation, authenticated transactions, and increased efficiency and trust between ecosystem participants [25–27].

Consequently, blockchain-based applications have found wide application in supply chain management. A research effort in this direction was already made in previous work [10,28–30], where an algorithmic optimization of the profit mechanisms in supply chain management is illustrated and implemented in the form of a SC to enable supply chains to keep balanced in terms of information and decision-making power while boosting business results similarly to vendor vertical integration systems managed by dominant corporate players. In other works, the security of data involved in the system is the crucial goal; for instance, Vo et al. [31] proposed a supply chain management model based on Hyperledger Fabric to reduce the problem of data explosion and prevent data manipulation and disclosure of sensitive information.

A plethora of evidence demonstrated the vantage brought by DLTs and blockchain in tracing, certifying processing stages, and offering trust in various fields, such as renewable energy and vaccine supply chains [32–34], and secure data sharing has been widely proven and discussed [35,36].

Several IoT applications rely on centralized architectures, which are characterized by the failure of a single point, trust management, and privacy concerns. Sharma et al. [37] conducted a recent review of blockchain-based applications. They reported that blockchain-based architecture for IoT connects all devices in a distributed manner and provides a secure method to share resources or data. Moreover, previous research showed that integrating IoT with the IOTA's Tangle can manage monitoring tasks on the agrifood supply chains, reducing resource consumption, thus improving the sustainability of cultivation [38,39].

### 3.2. Transformative potential and applications of Web3

Web3 represents the vision of a new, blockchain-based web that evolves into a read/write/own version of the Internet. In this paradigm, users participate in web communities and have a financial stake and greater control over them. Gan et al. [40] have comprehensively surveyed the evolution and potential impact of Web3 across various sectors, emphasizing its role in fostering a decentralized, intelligent, and user-centric web experience. This shift significantly transforms how information is created, distributed, and utilized across domains. Web3 ecosystems can leverage blockchains, cryptocurrencies, and decentralized finance (DeFi) to forge a more secure, transparent, and interoperable system in the banking and financial sector [41,42]. These technologies introduce decentralized ledgers, SCs, tokenization, and peer-to-peer lending, offering benefits such as trust, immutability, traceability, and interoperability. Moreover, Web3 challenges traditional banking paradigms by empowering users, reducing intermediaries, and minimizing transaction fees. The research and academic sector stands to gain from Web3 through the adoption of blockchains, decentralized networks, and SCs, enhancing the quality, integrity, and accessibility of scientific outputs like publications, data, and peer reviews. Web3 fosters novel collaboration forms, recognition, and rewards for researchers through decentralized autonomous organizations [43].

### 3.3. Decentralized identity management

Identity management refers to the policies and technologies that ensure only authorized individuals can access an organization's resources. The management of digital identities supplies an approach to authorized and identified users in every kind of system. Besides the traditional centralized or federated identity management model, the self-sovereign identity (SSI) model is among the most innovative. This model fits perfectly with the blockchain decentralized schema due to its bases on decentralized identities and verifiable credentials (VCs). A DID is created from a private/public key pair and is a unique user identifier. On the other hand, VC is a document digitally signed by the assigned to a DID [44].

SSI is a promising paradigm evolution for identity management; nonetheless, in a recent work, Hoops and Matthes [45] showed that SSI-blockchain-based implementations have a dependency on a particular type of blockchain, which is in contrast with the interoperability requirement. Therefore, SSI models must be further explored and developed to be considered an efficient improvement in the industrial sector.

### 3.4. Literature-extracted requirements

The literature review highlighted how blockchain-based systems, and thus BBIEs, must cope with a traceability property enhanced with transparency and immutability, likewise preventing information disclosure. Traced data must be certified in each processing stage, and trust must be guaranteed in a trustless environment given no central authority by leveraging distributed consensus.

Furthermore, the blockchain-based model must benefit companies and their interactions with customers and all the involved stakeholders. Another key property is interoperability. In fact, blockchains can give their properties in legacy to integrated systems. Therefore, blockchain-based systems must allow their component modules to integrate with each other and with external systems.

## 4. Method

In this section, we provide the experimental plan followed to reach the goal of our work. To gather the current needs of companies that operate adhering to B2B and B2C models, we define a first research question:

- **RQ1:** *What are the requirements to consider in designing a modular and scalable blockchain-based architecture to meet the authentication and data management needs of B2B and B2C organizations?*

To gain insight beyond what has been found in the literature, we set up a questionnaire targeted at professionals to bridge the gap between theory and practice. The requirements that emerged from reviewing the literature are devoted to dealing with specific scenarios, such as supply chain management, and the perspective is focused on problems, solutions, and related strategies, architectures, and systems. In this sense, what is missing are industrial needs, alongside those that are the crucial way to operate in the enterprise sector.

The results of the questionnaire are dedicated to addressing **RQ1** alongside knowledge collected through the literature review.

*4.1. Questionnaire*

We prepare a comprehensive questionnaire designed to gather insights from professionals across companies offering both B2B and B2C or citizen services. The goal of the questionnaire is to collect valuable data regarding the implementation and utilization of our proposed architecture within various organizational settings. Specifically, we aim to understand how these companies manage user data, typical challenges related to data management, data security, and interactions they encounter in their operations.

The questionnaire is strictly confidential, ensuring the privacy and anonymity of respondents. It commences with demographic questions to categorize respondents based on their employer, role, and experience level. Subsequently, it delves into three primary aspects:

1. **Authentication Systems**: We inquired about the authentication systems utilized by the company, including online authentication strategies such as:
   - *Multifactor authentication*, a security system that requires the use of more than one identity verification method, such as passwords, tokens, or fingerprints, to access an account or platform.
   - *Single sign-on*, an authentication system that allows users to access multiple applications or services with a single identification, simplifying the login process and improving the user experience.
   - *"In-house" authentication systems*, developed and managed directly by the company.
   - *Biometric identification systems*, based on biometric characteristics such as fingerprints, facial recognition, iris, retina, hand geometry, voice prints or biometric signatures.
   - *Identity-as-a-service systems*, a cloud solution that provides identity management services, including authentication, authorization, and access management. It allows companies to centrally manage user identity through a subscription-based platform.

   Understanding the authentication landscape allows us to assess the existing security measures in place and their effectiveness.

2. **Data Exchange and Interaction**: We explore how companies interact and exchange data with other entities, shedding light on collaborative processes and potential integration challenges.
   In particular, respondents are asked whether their company shares information about their business with other companies. This inquiry aims to gauge the extent to which companies engage in outbound data-sharing practices with external entities. If a company indicates that they engage in outbound information sharing, we further inquire whether other companies reciprocate by sharing their data with the respondent's company. This question seeks to assess the presence and nature of inbound data-sharing arrangements within the ecosystem.
   The inclusion of questions related to data exchange and interaction in the questionnaire provides valuable insights into the collaborative dynamics among ecosystem participants. Understanding the patterns and practices of data exchange and interaction between companies is crucial for identifying collaborative opportunities, addressing interoperability challenges, and enhancing data-driven decision-making processes. By soliciting insights into outbound and inbound data-sharing practices, we aim to gain a comprehensive understanding of the flow of information within the ecosystem and its implications for digital identity management and security.

3. **Legacy System Presence**: Additionally, we investigate the presence or absence of legacy systems within organizations, recognizing their potential impact on digital identity management and security practices. Respondents are asked whether their company currently utilizes legacy systems. If a company indicated the use of legacy systems, we further inquire whether these systems interface with new generation systems. This aims to understand the extent of integration between legacy and modern systems within organizational infrastructures. Finally, respondents are asked to provide their perspective on the potential for improvement in the interface between legacy and newer systems. Understanding stakeholders' perceptions of existing interfaces allows us to identify opportunities for enhancing interoperability and system efficiency.
   Assessing the presence and integration of legacy systems is crucial for understanding the technological landscape within organizations and identifying areas for improvement. Legacy systems, while serving critical functions, can pose challenges in terms of compatibility, security, and scalability.

Furthermore, the questionnaire includes inquiries about user profiling practices, ethical considerations, and data privacy concerns. We seek to understand respondents' perspectives on the ethical implications of user profiling and the types of data collected for such purposes. Open-ended questions are also included to solicit insights on privacy issues, security risks, and digital identity theft, facilitating a comprehensive understanding of the challenges faced in digital identity management. Digital identity theft [46] is a crime in which a malicious individual fraudulently acquires and uses another person's personal information on digital platforms, in order to commit scams, financial fraud, or other crimes.

In summary, the questionnaire was meticulously designed to address key aspects of digital identity management, security practices, and organizational dynamics within diverse business environments. We are confident that the insights gathered through this survey will contribute significantly to our research objectives and enhance our understanding of how our proposed architecture can address the evolving needs of modern enterprises.

*4.2. Questionnaire results*

We collected 11 responses from companies that are active in the ICT field, companies that provide products to the final consumer, a company that deals with earth observation, and a government organization. Regarding the role of the interviewees within their companies, five define themselves as "developers", three as "R&D directors", two as "IT architects" and one as a "managing partner". More than half of the respondents (54.6%) have 5 or more years of experience in the sector. Only one-fifth have less than 2 years.

The interview highlights how the vast majority of companies use online authentication systems. Only one respondent answers "no", whereas two others answer with "not applicable".

User profiling is the process of collecting and analyzing individual online personal data to create a detailed profile that predicts behavior, preferences, and characteristics. It is often used for advertising or targeted marketing purposes. However, we find that only two of the companies interviewed currently profile users (18%). However, 55% of the respondents believe that user profiling would be useful. All but one of the respondents who do not believe that profiling is useful also believe that profiling is not ethically correct. Approximately 73% of the respondents believe that privacy reasons must have an identity management system that does not store user data.

The main privacy and security issues and risks that our respondents believe arise in digital identity management are as follows:

- Identity theft, i.e. theft of user credentials outside the system, and consequent access to the system or to social profiles by malicious unauthorized persons.
- Inappropriate use of profiled data.

- Data breach, i.e., theft of data kept in the system and consequent disclosure of sensitive information, with loss of reputation on the part of the firm and risk of legal action for compensation.
- Loss of data, and consequent loss of company's assets and competitive advantage.
- The management and storage of data are costly and difficult due to the too stringent legislation.

Two-factor authentication is used by 45% of respondents, whereas 73% use single sign-on authentication. Only 18% of respondents use Identity as a Service (IDaaS) systems on the cloud.

There are 82% of respondents believe that bio-metric authentications are useful. The same percentage of respondents believes that there are several, or even many, cases of identity theft, but nobody has actually experienced it in their organization.

Every respondent reports that their company receives data from other companies, yet only 27% indicate that their own company reciprocates by sharing data. Despite 83% of respondents believing that data sharing among companies is efficient, both the literature and identified risks suggest otherwise.

Finally, 55% of the respondents use legacy systems, two-thirds declare that these systems are interfaced with new-generation systems, and 80% of them believe that this interface can be improved.

When answering **RQ1**, the requirements to be taken into account that emerged from both the literature and the questionnaire are as follows:

- Traceability: refers to the ability to accurately track and reconstruct the history, usage, item locations, processes, or events;
- Privacy: involves safeguarding sensitive information and controlling access to personal data to ensure confidentiality and prevent unauthorized use or disclosure;
- Resilience: refers to the capability of a computer system or network to maintain functionality and operational continuity despite disruptions, failures, or attacks, which is often achieved through redundancy, fault tolerance, and rapid recovery mechanisms;
- Secure identity management: involves the establishment and maintenance of trusted methods for authenticating and authorizing users;
- Integrity: ensures that data remain accurate, consistent, and unaltered throughout their lifecycle, safeguarding against tampering;
- Confidentiality: ensures that sensitive information is protected from unauthorized access, also maintaining the secrecy of data;
- Interoperability: refers to the ability of different computer systems, software applications, or components to seamlessly exchange and utilize data or services;
- Secure data sharing: involves the controlled and protected sharing of information between authorized parties;
- Data certification: assures data reliability and trustworthiness.

These requirements are carefully addressed in designing a modular and scalable blockchain-based architecture, to propose a solution that effectively meets companies' specific needs while promoting collaboration and innovation among different stakeholders.

*4.3. Grounded theory*

Our system's architecture is designed following the principles of grounded theory (GT) as provided by Glaser and Strauss [47]. GT is an inductive research methodology that focuses on generating theories directly from data, contrasting with the traditional hypothetico-deductive approach that tests pre-existing hypotheses [48].

To apply GT, we first review existing literature to understand the current research landscape. This review informs the development of our survey which consists of both closed and open-ended questions designed to gather qualitative data from professionals in the field. The insights gained from this survey are instrumental in addressing the real-world needs and components relevant to professionals, thereby grounding our architectural design in empirical data.

GT has been effectively applied across various disciplines such as software engineering, sociology, education, and management theory, noted for its robustness in generating practical and theory-driven insights [49].

To integrate the findings and identify how a BBIE can meet the elicited requirements, we pose a critical research question:

- **RQ2**: *How can a BBIE architecture address the collected requirements?*

This approach allows us to directly incorporate user feedback and current academic discourse into our system design, ensuring that the architecture addresses both theoretical and practical considerations effectively.

**5. Architecture for blockchain-based information ecosystem**

In this section, we discuss the architecture we devised to answer **RQ2**, that emerged from the results obtained in Section 4, taking into consideration of the requirements to be addressed that emerged from the literature as well. In particular, we introduce a specific component related to identity management, since this aspect emerged as a strong need from the companies addressed by the interviews. Aspects related to the state-of-the-art are discussed in Section 3; therefore, to begin with, we summarize the capabilities of blockchain-based systems, in addition to the necessities that arise as a result of the questionnaire.

As industrial business flows become increasingly complex, the sharing of business data and information among participants has become necessary. Currently, the leading company involved in the business flow manages the trust centrally. However, introducing blockchain technology as a trusted component can make data sharing transparent, reliable, and tamper-proof. This makes blockchain technology an ideal inter-enterprise data keeper. In fact, blockchain applications are widely used in commercial companies, such as supply chain management, healthcare, and collaborative economic optimization of businesses [5,28,50].

Although blockchain-based applications offer numerous advantages and innovative features, they have limitations when used in an industrial context that typically involves several interconnected modules. To overcome this limitation, our proposed solution integrates the blockchain-based application as an ecosystem module with the existing information system. The integrated blockchains with the information ecosystem create BBIE [8].

The proposed architecture is depicted in Fig. 1. It comprises both on-chain and off-chain components to enhance performance and functionality. On-chain activities provide security, transparency, and immutability, while off-chain solutions offer scalability and efficiency improvements for specific transactions and processes.

*5.1. Off-chain components*

All modules that belong to traditional information systems are considered entities outside the blockchain environment.

- *Storage*: The storage module of industrial ISs is an essential component that facilitates the collection, processing, and retrieval of vast amounts of data generated by industrial processes. This module comprises both hardware and software designed to store and manage data from various sources, such as human input, sensors, and automated processes. The rise of cloud computing and big data technologies has enabled the development of scalable storage solutions for modern industrial systems [51]. In a BBIE, it is essential to determine which data should be stored in off-chain and on-chain storage. Our approach is to base this decision on the need to share the data while ensuring tamper-proof transparency. Moreover, the traditional storage module is responsible for preserving intra-enterprise data, guaranteeing that information is isolated from other parties involved in the business. Large volumes of data, such as documents and photos, must also be kept off-chain. The document's integrity and date can be ensured by storing the hash digest of the document along with a link to retrieve it on the blockchain.
- *Business Layer*: Transactions can be expensive, so complex computations are often recommended off-chain rather than on-chain. In an off-chain context, off-chain business logic executes the internal operations of a company's information system. This means that computations are performed outside of the blockchain-based system, for example, on an intermediary service's server. On the other hand, in an on-chain scenario, computations are executed and verified within the blockchain-based system [52].
- *IoT Devices*: IoT devices are equipped with sensors that enable real-time data collection from industrial processes. This allows the capture and processing of large volumes of data. By passing these data through IOTA, and recording them on a blockchain, their integrity can be ensured, as they become tamper-resistant and have an immutable audit trail. SCs on the blockchain can be automatically triggered based on predefined conditions in IoT data. This means that IoT sensors can update the blockchain with real-time information about the movement and condition of goods. This information can then trigger SCs for automated payments or notifications. Therefore, IoT sensors play a crucial role in collecting and processing data from industrial processes and can enable automated SCs for various applications.
- *User Interface*: Interaction with the business layer is commonly facilitated through web or app interfaces, as well as dedicated application programming interface (API) layers. However, this is not explicitly indicated in the diagram.

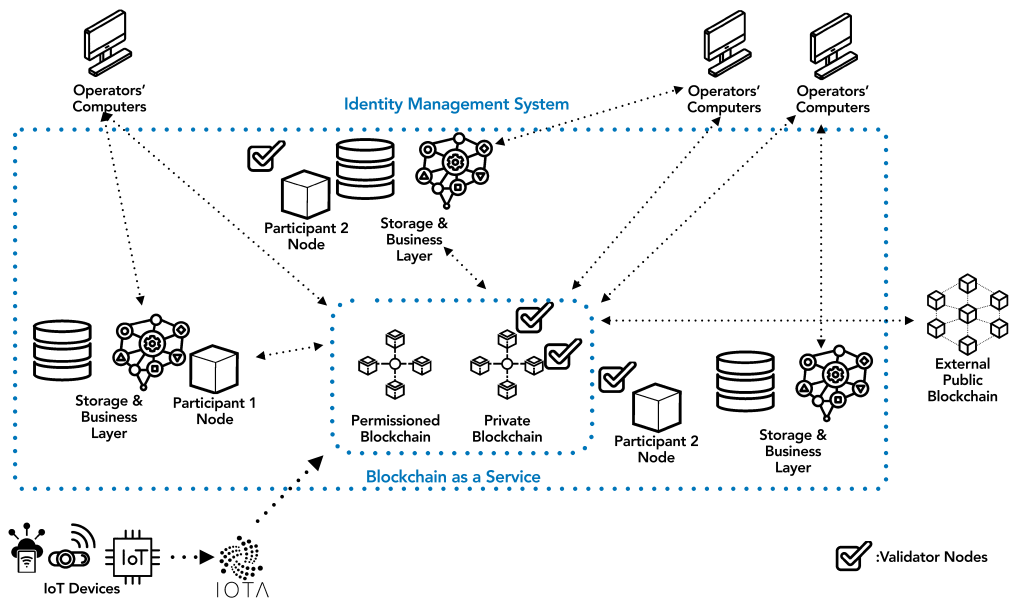

**Fig. 1.** Blockchain based information ecosystem architecture.

### 5.2. On-chain components

On-chain modules are part of the BBIE that performs actions or stores data directly on the blockchain network using its native features, including SCs and transactions.

Companies can either create their own blockchain or use a BaaS. In the scenario presented in Fig. 1, the latter option is shown, and the on-chain components are included in the BaaS component.

The BaaS provider uses a hybrid system of private and permissionless blockchains.

- *Private Blockchain*: is a network consisting of regular and validator nodes, which participant companies can run.
- *Permissioned Blockchain*: can either be accessible to everyone or restricted to authorized individuals only.
- *Shared Business Logic*: controlled via SCs execution.
- *An Explorer*: runs on one or more nodes that hold a copy of the blockchain. It allows users to browse the actual state of the blockchain without the mediation of the user interface, allows them to be used by anyone, and enhances transparency.

### 5.3. Other components

Apart from the on-chain and off-chain components, we also have other more components that do not clearly belong to any one of these categories. In fact, in our ecosystem architecture, the role in the blockchain for providing security, transparency, and immutability, is played by the BaaS component which may include different blockchain solutions, such as a permissioned and a private blockchain. All the remaining parts seem to belong to the off-chain components. Nevertheless, the IOTA component needs a specific classification. In fact for the interaction with IoT sensors and devices, the IOTA solution is more suitable. IOTA serves as a decentralized platform facilitating secure and cost-free transactions between IoT devices through its Tangle technology. By utilizing IOTA, IoT sensors can seamlessly communicate data in real-time while ensuring integrity and reliability. Subsequently, the validated data are securely anchored onto a blockchain for transparent sharing and immutability, establishing a robust foundation for trust and accountability in data management systems. In this sense, this is neither an off-chain component, since IOTA itself is a DLT and is often referred to as a blockchain, nor an on-chain component, since these are specifically outlined in the BaaS module.

Another part of our architecture that does not properly belong to the "on" or "off" chain components is a link to an external public blockchain. This component links the BaaS to an external public blockchain that can be used to record the hash digest of the last block mined locally. This process is called anchoring and it guarantees that the BBIE has the same level of immutability as a public blockchain. This anchoring can be done periodically, such as every 24 hours, or a specific event can trigger it.

Finally, we'd like to include the identity management system component in this section. In fact, as it will be made clear later on, we envision two different control levels on the access. The first one is ruled by a permissioning dAapp, namely, a set of SCs that can be deployed on the blockchain and rule the permissions at the global ledger level. Furthermore, we include identity management for the blockchain infrastructure itself, such as different access privileges for different kinds of nodes. At the same time, we devise actors who do not need to participate in keeping up and maintaining the infrastructure but only need to operate on the ledger through read and write operations; thus, the identity management module can be set to allow reading operations only on part of the ledger or on the entire ledger.

### 5.4. BBIE component interactions

BBIE components can execute their services based on specific needs. Interactions are event-reactive, such as HTTP calls and certain met conditions, including the time passed since the last anchoring, as discussed in Section 5.2. The BBIE modular services can be accessed in any order, allowing them to be tailored to the specific needs of different industries. To ensure clarity in the information ecosystem, off-chain business logic and BaaS services can be accessed by calling the API endpoints.

Public blockchain services, such as certification transactions, as well as private and permissioned block anchoring transactions, can be accessed using the application binary interface (ABI). Contract ABIs are crucial components for interacting with Ethereum SCs outside the blockchain and for contract-to-contract interactions [53]. External users use the ABI interface to retrieve information about the anchored state of data managed by BBIE.

When dealing with the information used to guarantee access and control over it in the public blockchain, BBIE only needs the public address to perform authentication. In detail, SCs use the mapping between addresses and roles to enable the permissions related to calling a given function, even in this case, no sensitive data are treated on the back of privacy guarantees.

Private BaaS is accessed by providing an OAuth2-compliant token, in this sense, to grant privacy is crucial to not include sensitive data. Hence the token must contain an ID in the shape of UUID or GUID, a role that identifies the set of operations that can be called, according to the granted privileges. When issuing a token, it is a convention to include the information of both the instant of issuance and expiration within the token. This helps to provide a complete and accurate record of the token's validity. Notice that SCs deployed on the permissioned blockchains can interact with those shared on the permissionless. To achieve authorization in this scenario, the permissioned blockchain must use an identity to call public SCs, which even consists of a wallet address in this case.

Other interactions, such as querying the database and interfacing with the business layer, are the same as those performed by traditional information systems.

### 5.5. Actors

The proposed architecture comprises four types of actors:

- *The Participants*: On the one hand, each participant manages and runs a private blockchain node. On the other hand, its storage and business layer are managed off-chain. For simplicity, we show only three participants, but there could be more.
- *The Validators*: As represented as checked boxes in Fig. 1, the validators are key participants in the ecosystem. They keep a copy of the blockchain and validate transactions through a consensus mechanism.
- *The Operators*: Each participant has its own operators who can communicate with their storage, business layer, and the blockchain.
- *The External Users*: They are the users who can only communicate with the blockchain in read-only mode using an explorer but may access authorized documents in participants' archives.

There may be a channel that allows participants to communicate directly, but it is not specified in the figure. The solution presented integrates a blockchain-based application with the information system as a module of the ecosystem. The blockchains depicted in the image are intentionally generic.

The BBIE model comprises two environments: an internal environment that interacts with the private network and an external environment that interacts with the public network. The private network includes private and permissioned blockchains, along with an off-chain business layer. Only individuals affiliated with the companies, involved in the BBIE, and authorized by the identity management system have access to this network. Fig. 2 illustrates this model. External users who are not affiliated with the companies involved in the shared business can access certified data and traceability information by interacting with the public network. This network provides data stored on the public blockchain(s). Although integrating different technologies to enable interoperability may be challenging, BBIE can overcome these challenges by utilizing communication standards such as REST APIs and ABI interfaces to integrate heterogeneous components.

As shown in Fig. 2, the BBIE architecture needs to manage the identities of various actors involved. In general, there are two different issues: identity management for on-chain access and identity management for storage and business layer access.

The latter can be dealt with using the usual technologies for DB management, and since they are well known, we do not include this part in our discussion. The former is more delicate for different reasons. First of all, data written on-chain cannot be deleted or changed afterwards, different from traditional DBs. Thus, it is not possible to remedy and correct potential errors or mistakes, either unwanted or fraudulent. Second, identity management should be compliant with data protection regulations, such as GDPR, since once sensitive data are recorded on-chain, they are exposed forever and the right to be forgotten cannot be implemented. Third, if one actor has granted access to a node, it can access all the data on the chain.

Without losing generality, we describe identity management in a permissioned blockchain since it is representative of all the other cases and is also the most common one in BaaS. Identity management in a permissioned blockchain, where participants are known and trusted, is still a critical aspect, albeit with some differences from public, permissionless blockchains. In a permissioned blockchain, the network is restricted to a specific set of participants who are known and authorized to participate.

We devise two different aspects. The first is about the identity management for the blockchain infrastructure itself. In fact, in permissioned blockchain, participants in the infrastructure management have different roles, and these roles define what they are allowed to do within the network. For example, in the first level, which can be almost thought of as a physical level for the network infrastructure, we can consider simple nodes or validator nodes. The former can keep a copy of the entire ledger, provide access points to the ledger to external users, and send transaction requests or act as a proxy for requesting transactions. The latter also entitled, in addition, to validate transactions and blocks and can update the state of the ledger.

In the second level, which can be thought of as a purer logical level, we can consider addresses with different privileges. These may be entitled to act on the first level so that they can decide or approve new nodes joining to the blockchain network or can decide to upgrade a simple node to the role of validator.

The second refers to the actors who need to use the infrastructure for reading and writing operations. These actors need to be grouped according to their roles and set their permissions and privileges according to their roles. These actors do not need to participate in keeping up and maintaining the infrastructure, although they could also do this in principle, but only need to operate on the ledger through reading and writing operations. Within this categorization, the identity management module can be set to allow reading operations only on part of the ledger

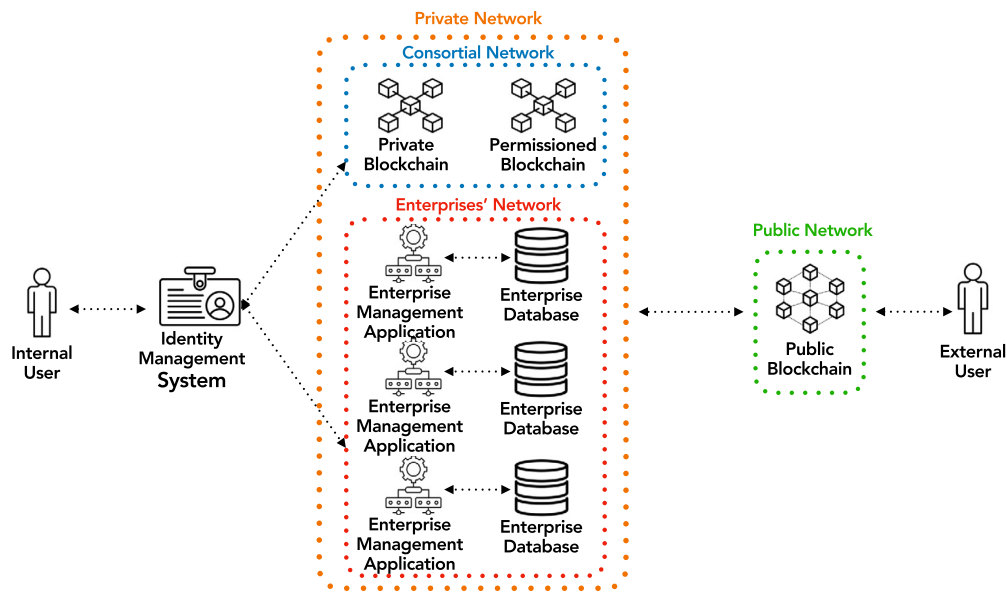

**Fig. 2.** Blockchain-based information ecosystem model.

or on the entire ledger and to allow writing operations on part or on the whole ledger as well.

Here, we envision two different control levels on the access. The first one is ruled by a permissioning dApp, namely, a set of SCs that can be deployed on the blockchain and rule the permissions at the global ledger level. This control level's paradigm is the one Consensys provides on the Hyperledger-Besu blockchain infrastructure.[1] The second one is obtained as usual by deploying dApp(s) on the blockchain which provide access control policies to the data they manage and have access to.

In the first case, the dApp has global control on the blockchain and is able to set a pool of administrator users, identified through their public address and managed by the usual private-public key pair, with overall privileges. An administrator can decide to enroll other nodes on the blockchain, decide which addresses are entitled to SC deploying, and also which addresses can be enrolled as administrators.

In the second case, which is the standard in public blockchains, SCs can be deployed on the ledger and can be set with different privileges and access rights for the owner (the actor who deploys the set of SCs) and for other users, with policies decided by mean of public or private methods or variables in the SCs and by mean of address identification.

In summary, the architecture that we devised allows for the satisfaction of the requirements given as the output of the conducted GT; therefore, we discuss the way that is accomplished before evaluating our BBIE architecture against a case study. Our architecture enables privacy granting by leveraging the proposed identity management system, it also relies on the properties of the DLT involved, in detail. In the on-chain world, identities are shaped into addresses, which are public and completely missing user-related data.

Resilience is implied by the decentralized and distributed nature of the ledgers. Indeed, given the non-centralized schema, the BBIE architecture can rely on the no-single-point-of-failure granted by the distributed ledger.

The security of identity management is achieved by minimizing the user data in the off-chain world through the OAuth-based encoded tokens and the public key derived by the private key generated in the wallet creation stage when dealing with the blockchains. In this sense, this management model also ensures the confidentiality property; only those with the role and the privileges to access the BBIE resources can pass the authorization layer of the IMS.

Data integrity is guaranteed by the use of the blockchain's immutability property. BBIE architecture grants interoperability to the communicating components and external modules exploiting services exposed through API and ABI, and formatting responses into standardized object notations.

Secure data sharing, traceability, and data certification are enabled by the capabilities of distributed ledger technologies, as outlined in Section 3.

**6. Case study**

In the following, we will illustrate a case study crafted by considering the GT-arisen requirements. The goal of such a case study is to evaluate whether our architecture addresses the gathered needs. The case study analyzed deals with the management and monitoring of an agri-food supply chain. The agri-food companies work together to share the numerous growing, harvesting, and processing stages needed to bring a finished product to the final customers.

The needs of this cooperative business include the sharing of data between parties, real-time monitoring of measures related to the entire supply chain of the goods, and the implementation of a reliable traceability system to guarantee food quality and safety. On the other hand, identity management is a paramount aspect in both tracing the identity of involved parties and authorizing and authenticating users.

Thus, we begin our case study by taking care of the requirements addressed by real-time monitoring, data sharing, and the traceability system enabled in a BBIE. In detail, real-time monitoring and trusted traceability have become crucial in ensuring food safety, quality, and sustainability in the growing and production processes. DLT and IoT, when used together, are cost-effective solutions that meet all requirements. In fact, IoT devices can obtain environmental and product-related measures; the DLT can collect these measurements by sharing them with those involved in the agri-food supply chain.

First, we clarify the role of the three types of DLTs involved in the given case study.

1. **DAG-based DLT**: IOTA, chosen as a DAG-based DLT due to its ability to act as a mediating link between a vast number of IoT sensors and the BBIE, takes care of processing in real-time the stream of data received from the IoT sensors devoted to collecting environmental measures and food features. Having no transaction costs, IOTA stands out as an efficient gathering point as well as extremely low latency, thanks to the absence of consensus algorithms, allowing for in-site sustainable monitoring. IOTA collects data received through streams coming out from IoT devices leveraging several protocols, such as Message Queuing Telemetry Transport (MQTT), Constrained Application Protocol (CoAP), HTTP, or WebSocket. For this case study, the MQTT protocol is a good fit, as it is designed to be efficient and lightweight, making it suitable for devices with limited processing power and memory resources.

---

[1] https://github.com/Consensys/permissioning-smart-contracts/tree/main.

2. **Permissioned blockchain**: Hyperledger Fabric (HLF) is employed to certify information among the parties involved in the consortium, it is in charge of sharing validated data regarding the production process and solving the visibility problem. In addition, it takes care of sharing information about the certified data collected in IOTA, enhancing the security and trustability of such data among collaborative parties.

3. **Public blockchain**: Ethereum allows BBIEs to ensure that everyone, including end customers, can know the material information, such as the provenance and other key characteristics of the final product, as well as the steps followed in production, providing the widest possible transparency.

While IOTA is connected with the IoT sensors with the MQTT protocol, blockchains supply their services with the BaaS; the BaaS provider utilizes a combination of private and public blockchains to ensure the authenticity of data pertaining to inner consortium companies' progress on a private blockchain, as well as data related to the completion of entire processes on both public and private blockchains.

Another aspect requiring interoperability is the communication between IOTA and the on-chain business logic. In the context of a BBIE, SCs are implemented and executed on the permissioned blockchain, incorporating specific logic to interact with the IOTA network, which is enabled by specific libraries such as IOTA.go[2]. This demonstrates the integrability of BBIEs, bringing an open road to the development of dApps that can leverage the main features of the two kinds of DLTs.

This hybrid system provides a reliable and secure mechanism for verifying information accuracy while allowing for flexible data sharing across multiple parties. Its deployment enables organizations to ensure the trustworthiness of shared information among stakeholders. By leveraging blockchain-based features, the hybrid BaaS guarantees the confidentiality of business information that remains within the consortium. Simultaneously, it supplies public access to traceability information for customers, also ascertains the traceability and immutability of data. Furthermore, in the previous steps, data collected via IoT sensors are managed by IOTA, which can handle and execute transactions more efficiently than chain-based DLTs, fitting optimally in an IoT scenario characterized by a vast volume of transactions, such as food supply chain monitoring.

Identity must be managed across the flows among the listed components. Thus, we illustrate the identity management scheme by means of a use case in which we identified five actors in a BaaS, where the covered need is the management of an agri-food supply chain. These are two agri-food companies with different employees, two certification agencies with their employees, and one set of BaaS providers that manage the ledger and provide blockchain services.

A further actor could be included as a blockchain software provider, namely, a software company specialized in writing specifically tuned dApps for customers' needs. In our use case, the two agri-food companies can contact the software/dApps provider for the deployment and configurations of the system for managing the supply chain. The role of the dApp provider/developer can actually be played by one or more participants in the blockchain infrastructure and be included in the BaaS global system.

---

[2] https://github.com/iotaledger/iota.go

**Table 1**
Mapping of case study components with BBIE components.

| Case Study Component | BBIE Component |
| --- | --- |
| Hyperledger | Permissioned blockchain |
| Ethereum | External public blockchain |
| IOTA | IOTA |
| Sensors and monitoring devices | IoT devices |
| DMS, DBMS, and private custom software | Off-chain business and storage layer |
| Companies and authorities | Partecipants |
| Customers | External user able to explore the public blockchain |

In the following, we assume this latter configuration where the BaaS also includes the dApp specifically developed for the two companies. Since the use case also includes two public body actors, i.e., the certification agencies, it is straightforward that the dApp developed takes into account these roles.

In particular, the actors involved are:

- A company for the production of wine, responsible for cultivating, harvesting, fermenting, aging, and bottling wine. The company manages the winemaking process and needs cork stoppers for the bottling phase.
- A company for the production of cork stoppers that specilazes in manufacturing closures specifically for wine bottles. It ensures quality and compatibility with wine storage and aging requirements.
- A public administration in charge of sanitary and health verifications, such as the Italian ASL (Azienda Sanitaria Locale, which translates to local health authority). It plays a crucial role in ensuring the safety and quality of food products within the agri-food supply chain.
- Another public administration in charge of rules and compliance verifications and quality certification, such as the Sardinian Laore. By certifying and promoting the quality and authenticity of products, such as traditional food products with protected designation of origin (PDO) or protected geographical indication (PGI) status, it helps to enhance the reputation and marketability of the products.
- The set of BaaS providers. A pool of bodies, which includes the ASL (has the right to keep all the ledger for sanitary reasons), private companies specializing in blockchain technologies, and other public administrations that may grant the blockchain is provided to every customer with no discrimination or favors.

All actors perform steps along the chain and involve employees with different roles and permissions to be managed.

Fig. 3 displays the described supply chain. Each processing operation in a single processing stage is traced and certified on the consortial blockchain, gathering data dedicated to monitoring and quality guaranteeing from IoT devices passing through the IOTA DAG-based DLT. After each processing phase, information is permanently recorded on the permissionless blockchain, also granting access to such data to the final customers. Enterprises gain visibility of the processing state in each phase of the product's lifecycle, by obtaining certified insights through data recorded on blockchains in the consortial network. The same visibility is guaranteed to the authorities that oversee the goods' conformity, thus enabling the respect of compliance policies and regulations.

Each enterprise involved in the collaborative business has its own business logic and storage layer, dedicated to performing internal operations and data that are not meant to be shared. The identity managed system manages all identities of the entities cooperating according to business-defined policies.

Table 1 maps each component of the case study to the corresponding BBIE component, as shown in the general architecture in Fig. 1.

We discuss how BBIE addresses the requirements that emerged, thus responding to **RQ2**; in detail, for each need, it is underscored as follows:

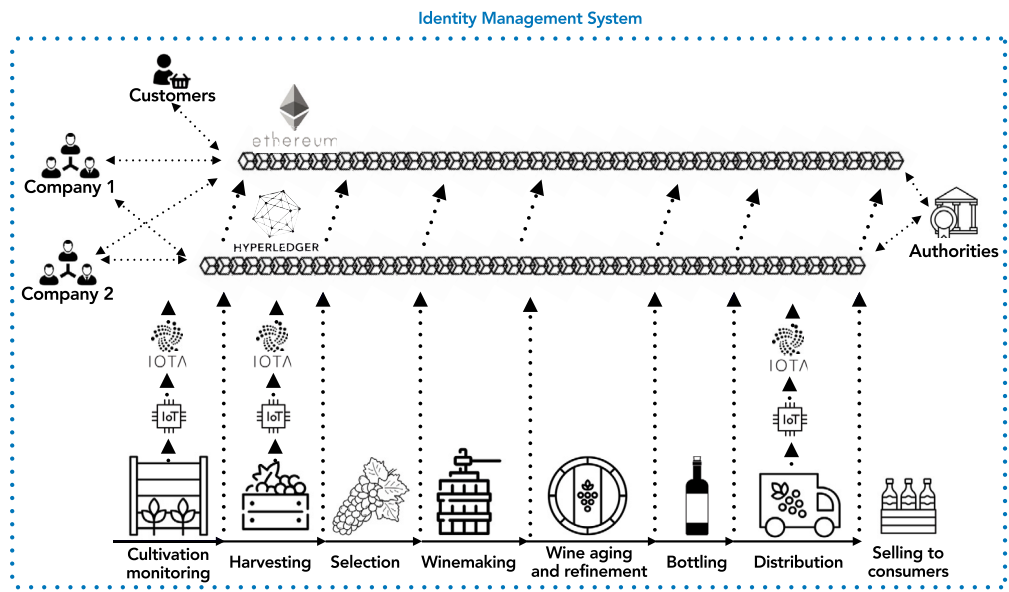

**Fig. 3.** The overall supply chain of the discussed case study.

- Traceability: Achieved through real-time monitoring and certified storage of individual processes. In simple terms, the IoT captures current state data and transfers it to IOTA. This information is then stored via transactions on distributed ledgers using SCs. Initially, this is done on the permissioned network, and subsequently, at each supply chain step or after 24 hours, it is transferred to the public blockchain.
- Privacy: Granted using authentication a privateness, only data that are meant to be shared are accessible in a public way.
- Resilience: Guaranteed through the no-single-point-of-failure guaranteed by DLT, in addition, data are redundant on two different blockchains of different natures.
- Secure identity management: Identities are treated following common and secure protocols and behaviors all along the component of the BBIE. OAuth-based identity management is shared among off-chain and permissioned blockchain. On the other hand, addresses serve as public blockchain identities. Moreover, privileges are permitted on the back of a given role, and permissioned nodes are under the authorization and certification policies of the private blockchain.
- Integrity: It is inherited from blockchain properties, once a piece of data is stored, it becomes immutable.
- Confidentiality: The hybrid managing strategy of the involved intensities enables keeping information confidential and accessible only to those authorized to view it.
- Interoperability: The proposed BaaS offers several advantages, including high integrability through the exposure of BaaS. This is achieved by using common protocols and notations like HTTP and JSON. Similarly, the off-chain business logic of a BBIE facilitates logic integration by providing interfaces that adhere to the above standards. While authenticating identities across different modules using various technologies may pose some challenges, leveraging the authentication flow highlighted in interoperability guarantees authentication even in the authentication processes.
- Secure data sharing: All private data are accessed by authorization only. As for the off-chain world, the OAuth2 protocol is used to manage authorized identities. Data in the private blockchain are accessed exclusively by authorized network participants or identities authorized to request it through BaaS. These data are shared on the same distributed ledgers.
- Data certification: Immutability, distributed consensus, and transaction transparency make data on the blockchain certified, as they are reliable, unaltered, and verifiable by any interested party.

> **Research Questions Summary**
>
> We report the posed research questions and briefly summarize the obtained answers.
>
> - **RQ1:** What are the requirements to consider in designing a modular and scalable blockchain-based architecture to meet the authentication and data management needs of B2B and B2C organizations?
> - **RQ2:** How can a BBIE architecture address the collected requirements?
>
> To sum up the answer to the first research question, we identifie, through reviewing the literature and analyzing the result of the questionnaire, nine distinct requirements, encompassing traceability, privacy, resilience, secure identity management, integrity, confidentiality, interoperability, secure data exchange, and data certification.
> Hence, regarding **RQ2**, a BBIE satisfies the requirements by utilizing the properties of its components. For instance, it achieves data certification, integrity, traceability, and secure data sharing by leveraging blockchain technology. The BBIE also ensures reliable identity management through an authentication protocol and enables identity sharing among modules that require privileges.

In summary, the companies involved in the supply chain could benefit from an improvement in real-time monitoring of their products, supported by data sharing and security through the blockchain ledger. In addition, identities are managed in two different manners, one is infrastructural, given by the authorized nature of permissioned blockchains, and the other one is logic-related, thoroughly addressing the requirements of proper access control.

## 7. Discussion

The core objective of an information system is to provide accurate, secure, easily accessible, and relevant information to aid organi-zations in decision-making and goal achievement. Essential features for such systems include traceability, authentication protocols, and analytics tools. This discussion outlines the industrial adoption of BBIE and highlights their advantages over traditional systems in meeting these requirements.

*7.1. Managing identities in BBIEs*

Identity management systems are crucial for organizations to securely manage access to resources, thereby mitigating security risks and streamlining identity management [54].

Blockchain technology introduces a decentralized and immutable ledger for identity data, reducing the dependence on central authorities. However, it complements rather than replaces traditional identity systems. It enhances security, transparency, and trust when integrated with systems like OAuth2, a protocol for authorization that allows authentication across applications without sharing login credentials [55] (see Section 6).

In BBIEs, OAuth2 tokens are crucial for managing identities. These tokens can be transmitted through business logic and BaaS to private or permissioned blockchains, which offer advanced security features like encryption and access controls. SCs on the blockchain verify identity and manage access, ensuring compliance with enterprise policies. Linking private blockchain blocks to public ones securely manages and verifies identities.

Moreover, direct interactions between blockchain services requesters and the blockchain can define access controls. For example, Ethereum SCs allow only authorized addresses, managed by contract owners, to execute specific actions. While self-sovereign identity (SSI) models show promise for identity management in BBIEs, they currently lack interoperability and standardization.

*7.2. BBIE integration with legacy or pre-existing systems*

Blockchain applications can leverage the BaaS concept to share infrastructure. SCs, accessible via the BaaS service interface, can be customized to integrate seamlessly with various systems, regardless of their architecture. Both service-oriented architecture and microservice architecture, which are commonly used in enterprise systems, facilitate this integration through standard communication protocols like REST APIs [56].

*7.3. Advantages of blockchain integration with enterprise information ecosystem*

Integrating blockchain technology provides shared control over information, ensuring a transparent and equitable distribution of rights, profits, and control. This decentralized approach allows multiple organizations to share and execute consortial business logic through SCs. All network nodes transparently execute and read these contracts and the ledger state, enhancing visibility and trust.

The advantages of blockchain for enterprise ecosystems include:

1. Entities like front-ends and devices interact with SCs on private Baas platforms.
2. Upon reaching a consensus, these contracts execute operations and interact with distributed ledgers, promoting transparency among business parties.
3. The most recent block of a private or permissioned blockchain is periodically anchored on a public blockchain or triggered by specific events, using ABI interfaces for public data access [57].

## 8. Conclusion and future work

This study has introduced a groundbreaking approach to redefining the management of information ecosystems through blockchain technology, significantly challenging the traditional centralized control exemplified by major corporate entities. Our concept of BBIEs leverages BaaS to seamlessly integrate blockchain technologies into diverse information ecosystems, proposing a novel architectural paradigm.

Our research delineates the flexible nature of BBIEs, which are capable of integrating a variety of new functionalities as they emerge. By leveraging the inherent trust, decentralized management of data, and shared business logic capabilities of blockchain, BBIEs facilitate a comprehensive reorganization of traditional business processes. This reorganization includes the deployment of distributed identity management systems, the secure recording of transactional data, and the implementation of advanced security protocols such as OAuth2. Moreover, the BaaS component bridges the gaps within legacy systems, addressing enduring challenges like the legacy dilemma and fostering a collaborative environment where information and business logic control are democratized.

Significantly, the introduction of BBIEs marks a paradigm shift in the application of blockchain technology, moving beyond typical use cases such as transaction traceability, to enable innovative business models across various industrial sectors. This capability allows BBIEs to support the development of new, multi-organizational business models, thereby catalyzing the transformation of traditional business operations into more transparent, efficient, and accountable practices.

Despite these advancements, the development of BBIEs is still nascent. Several avenues for further research and development are crucial for optimizing this framework and ensuring its effective integration and functionality within the global business infrastructure:

1. **Interoperability and Standardization**: Future studies should focus on enhancing the interoperability between different blockchain platforms and existing IT infrastructures to facilitate smoother integration and broader adoption.
2. **Scalability Solutions**: As BBIEs expand, scalability becomes a critical factor. Research into more efficient consensus mechanisms and off-chain transaction handling could provide necessary improvements.
3. **Regulatory and Ethical Considerations**: The decentralized nature of blockchain raises significant regulatory and ethical questions that need to be addressed to ensure that BBIEs operate within legal frameworks and maintain ethical standards.
4. **Sector-Specific Adaptations**: Tailoring BBIE architectures to meet the specific needs of various industrial sectors could provide practical insights and drive adoption in fields such as healthcare, finance, and manufacturing.
5. **Advanced Security Measures**: While blockchain offers enhanced security features, the evolving nature of cyber threats necessitates ongoing research into advanced security measures to protect sensitive data within BBIEs.

In conclusion, as we continue to advance and refine BBIEs, our aim is to lead their significant contribution to the evolutionary trajectory of information systems. By doing so, we anticipate that BBIEs will not only enhance the efficiency and security of business operations but also empower them with unprecedented levels of transparency and participant control, heralding a new era in the digital economy.

**CRediT authorship contribution statement**

**Francesco Salzano:** Writing – review & editing, Writing – original draft, Methodology, Investigation, Conceptualization. **Lodovica Marchesi:** Writing – review & editing, Writing – original draft, Methodology, Investigation, Conceptualization. **Remo Pareschi:** Writing – review & editing, Supervision, Methodology, Investigation, Conceptualization. **Roberto Tonelli:** Writing – review & editing, Supervision, Methodology, Investigation, Conceptualization.

**Declaration of competing interest**

The authors declare that they have no known competing financial interests or personal relationships that could have appeared to influence the work reported in this paper.

**Funding**

We acknowledge financial support under the National Recovery and Resilience Plan (NRRP), Mission 4 Component 2 Investment 1.5 - Call for tender No. 3277 published on December 30, 2021 by the Italian Ministry of University and Research (MUR) funded by the European Union – NextGenerationEU. Project Code ECS0000038 – Project Title eINS Ecosystem of Innovation for Next Generation Sardinia – CUP F53C22000430001- Grant Assignment Decree No. 1056 adopted on June 23, 2022 by the Italian Ministry of University and Research (MUR), F53C22000430001 the HALO (Hazard-Analysis and Anti-Counterfeiting Ledger Oriented Protection) project, funded by Sardegna Ricerche, CUP: F23C23000310008, and the W.E. B.E.S.T. (Wine EVOO Blockchain Et Smart ContracT) PRIN 2020 project, financed by the Italian Ministry of University and Research (MUR), CUP: F73C22000430001.